# Guiding, bending, and splitting of coupled defect surface modes in a surface-wave photonic crystal

Zhen Gao[1], Fei Gao[1], Baile Zhang*[1,2]

[1]Division of Physics and Applied Physics, School of Physical and Mathematical Sciences, Nanyang Technological University, Singapore 637371, Singapore
[2]Centre for Disruptive Photonic Technologies, Nanyang Technological University, Singapore 637371, Singapore
*Authors to whom correspondence should be addressed; E-mail: blzhang@ntu.edu.sg (B. Zhang)

## Abstract

We experimentally demonstrate a type of waveguiding mechanism for coupled surface-wave defect modes in a surface-wave photonic crystal. Unlike conventional spoof surface plasmon waveguides, waveguiding of coupled surface-wave defect modes is achieved through weak coupling between tightly localized defect cavities in an otherwise gapped surface-wave photonic crystal, as a classical wave analogue of tight-binding electronic wavefunctions in solid state lattices. Wave patterns associated with the high transmission of coupled defect surface modes are directly mapped with a near-field microwave scanning probe for various structures including a straight waveguide, a sharp corner and a T-shaped splitter. These results may find use in the design of integrated surface-wave devices with suppressed crosstalk.

Photonic crystals, also known as photonic bandgap (PBG) materials, are artificial periodic dielectric or metallic structures which can forbid the propagation of electromagnetic (EM) waves in a certain frequency range in all directions[1-3]. By locally breaking the translational symmetry of photonic crystal, highly localized point defect modes within the photonic bandgap can be created, being analogous to the localized impurity states in a semiconductor[4]. Thus only evanescent waves can penetrate, from this defect location, into the gapped photonic crystal within a finite distance, opening the opportunity for photons to hop from one defect cavity to the neighboring one. Following this concept, waveguiding along the impurity chains in photonic insulators[5] and through coupled defect modes[6] are theoretically investigated and experimental verified[7-9] in the context of photonic crystals. However, because photonic crystals work with Bragg interference, the mode profile of these defect modes are generally diffraction-limited, i.e. being limited to the scale of about half a wavelength.

On the other hand, electromagnetic modes supported on periodically textured metal surfaces, which are commonly termed as spoof (or designer) surface plasmons[10-22], possess spatial scales typically much smaller than the wavelength. In particular, some structured metal surfaces exhibit a complete photonic bandgap where no surface guided modes are permitted[23-26]. Recently, by introducing an engineered defect on a perfectly structured metal surface, a tightly localized subwavelength surface defect mode that emerges in the photonic bandgap is proposed theoretically[27-28] and verified experimentally[29] in the microwave regime.

Here we demonstrate that it is possible to merge the subwavelength feature of spoof surface plasmons with the waveguiding mechanism of coupled defect modes in the context of photonic crystals. This can provide a alternative way to manipulate the propagation of surface waves at subwavelenth scales on structured metal surfaces.

Compared with conventional waveguides of spoof surface plasmons[10-22] which generally suffer from scattering and crosstalk when packed densely, this coupled-defect surface waveguide (CDSW) is achieved through weak coupling between otherwise tightly localized surface defect cavities, which allows shaping the flow of surface waves almost as will[30] with minimal crosstalk.

We start our demonstration with a perfectly structured metal surface as an ideal surface-wave photonic crystal, as shown in Fig. 1(a). This surface-wave photonic crystal consists of a square array ($25 \times 25$) of cylindrical aluminum pillars with radius $r = 1.25$ mm, height $h = 5$ mm, and period $d = 5$ mm. By performing three-dimensional (3D) Finite Integration Technique (FIT) eigenmode simulations, we can obtain the band structure of the corresponding infinite surface-wave photonic crystal, which reveals a surface-wave bandgap from 12.6 GHz to 27 GHz, as shown in Fig. 1(b).

Our goal is to use this structured metal surface to control the propagation of surface waves through coupled surface defect cavities. We start by designing the most basic component: a surface defect cavity. For this purpose, we partially reduce the height of one pillar at the center from $h = 5$ mm to $h_d = 4.15$ mm, while keeping the rest pillars unchanged, as indicated by a blank dot at the center of the inset in Fig. 2(b). Thus the resonance frequency of the shortened pillar falls within the bandgap of surface-wave photonic crystal, and a defect state can be expected near the shortened pillar. To experimentally demonstrate this surface defect cavity, we employ a near-field scanning system working in microwave regime. The experimental setup consists of a vector network analyzer (R&S ZVL-13) and a pair of homemade electrically short monopole antennas placed in the near-field of the metal surface to excite and probe the surface waves, as shown in the inset of Fig. 2(a). The monopole source and probe are arranged in the direction normal to the sample surface and mainly sensitive to the vertical component of electric field (E$_z$). The probe antenna is mounted on a *xy*-motorized stage to measure the transmission between the two ports (widely called "S-parameter S$_{21}$") at any given position.

We then test this surface defect cavity. We first measure the transmission S$_{21}$ in the full surface-wave photonic crystal without any defect by placing the monopole source at the center of the surface-wave photonic crystal and the probe 1 mm above one chosen pillar, as shown in the inset of Fig. 2(a). Below the cutoff frequency of 12.6 GHz, we can see some weak resonant peaks in the transmission spectrum [black line in Fig. 2(a)], which means that some surface modes can be supported on this surface-wave photonic crystal. Then we replace the chosen pillar by a shorter one with height $h_d = 4.15$ mm and keep other experimental setup unchanged. We plot the transmission in presence of the single surface defect cavity in Fig. 2(a) with a red curve. Evidently, except the weak resonant peaks below the cutoff frequency of 12.6 GHz, an ultra-strong resonant peak appears in the bandgap of the surface-wave photonic crystal, which corresponds to the defect cavity created on the textured metal surface. Of equal interest is the spatial extension of the surface cavity mode created by the defect. Using the microwave near-field scanning stage, we measure the spatial distribution of the electric field (E$_z$) around the defect cavity at its resonant frequency 13.5 GHz, as shown in Fig.2 (b). It is clear that a highly localized spoof surface plasmon cavity mode is formed on the structured metal surface. For comparison, the electric field distribution of the defect cavity mode on a transverse *xy*-plane 1 mm above the textured metal surface obtained by FIT simulation is shown in Fig. 2(c). The agreement between the experiment and simulation results is evident. We also present the simulated Ez field distribution in the vertical central *xz*-plane of the single cavity, as shown in the inset of Fig. 2(c), which further confirms that the surface defect cavity modes are highly confined in the defect site and decay exponentially in both horizontal and vertical directions.

We now move on to more complex components for guiding surface waves. We first construct a straight waveguide by shortening a row of metallic pillars alternately, as shown in the inset of Fig. 3(a). The dispersion relation of the surface-wave defect modes can be obtained by applying the Floquet-Bloch theorem and using a commercial finite-integration time-domain algorithm, as plotted in Fig. 3(a). Here, even though the surface waves are guided by metallic pillars similarly to the case of spoof surface plasmons supported by a square array of metallic pillars[25,29,31], the dispersion relation is very different from that of conventional spoof surface plasmons, which starts at the light line and tends to zero group velocity at the band edge. Indeed, when excited, the cavity modes of each surface defect are tightly confined at the defect site and only a small portion of the field penetrates in the form of evanescent waves to reach the nearest neighbor. Thus surface waves can propagate via hopping due to the weak coupling between neighboring cavities. This nearest-neighbor coupling is exactly the classic wave analogue of tight-binding (TB) limit in solid state physics[32]. Hence, similar to coupled defect modes in photonic crystal[7] and coupled resonator optical waveguide (CROW)[6], the dispersion relation of CDSW exhibits a shape of sine function centered at the resonance frequency of a single surface defect cavity, as shown in Fig. 3(a), rather than a polaritonic dispersion relation.

We then construct and measure the normalized transmission spectrum through a straight CDSW with twelve unit cells, as shown in Fig. 3(b). We observe, within the photonic bandgap, a waveguiding band extending from 13.35 GHz to 13.65 GHz, with a normalized transmission of 80% and a band width of $\Delta\omega = 0.3$ GHz. We also plot the normalized transmission spectrum of the full surface-wave photonic crystal for comparison, which shows a wide forbidden band that starts from 12.6 GHz, being consistent with the band structure shown in Fig. 1(b). Measured and simulated maps of the spatial distributions of electric field 1 mm above the sample are presented in Fig. 3(c) and Fig. 3(d), respectively. Evidently, surface waves hop from a cavity to another, forming a very efficient and subwavelength confined propagation. In all the numerical calculations, as we were assuming ideal PEC boundary conditions, the modal propagation length, $l = [2Im(k)]^{-1}$, is infinite. However, in experiment we observed the decay of electric field [Fig. 3(c)] along the propagation direction due to the absorption (ohmic) losses of aluminum. One way to estimate the propagation length $l$ in a real CDSW waveguide operating at microwave frequencies is by measuring the decay of electric field in a straight CDSW, then the mode propagation length can be obtained by fitting the exponentially decay of the electric field amplitude along the waveguide. In the experiment, we have fixed a frequency of 13.5 GHz (the center of the normalized transmission spectrum in Fig. 3(b)) and the whole straight waveguide length is 11d (110 mm) with 80% transmission. Thus we can evaluate the propagation length of CDSW at 13.5 GHz ($\lambda = 22.2$ mm) is about 11.2 $\lambda$.

Since the propagation of surface waves in the proposed CDSW stems from the resonant nature of the defect cavity, we can tailor unit cells locally to shape the flow of surface waves along a prescribed path. We take the bending of surface waves through a sharp corner [inset of Fig. 4(a)] as an example. The measured normalized transmission spectrum through this sharp corner is shown in Fig. 4(a). A waveguiding band extending from 13.35 GHz to 13.65 GHz is observed, whose frequency range and normalized transmission are almost the same with the straight waveguide. The measured and simulated electric field distributions in a transverse *xy*-plane 1 mm above the metal surface are presented in Fig. 4(b) and Fig. 4(c), respectively. To further confirm that the sharp bends do not form cavities and are truly reflection free, we simulate a 180 degree

CDSW which consists of two sharp corners and present the Ez field distribution at 13.5 GHz in a transverse *xy*-plane 1 mm above the metal surface, as show in Fig. 4(d). Evidently, the transmission of surface electromagnetic waves across multiple sharp corners of CDSW is possible and almost perfect without reflection and scattering. Indeed, there are mainly three different mechanisms that guarantee surface waves can be perfectly routed through this CDSW sharp corner. First, according to Yariv's CROW thery[6], if a resonant cavity mode owns four-fold rotational symmetry, it is evident from symmetry considerations that one can make a perfect 90º bend, since the coupling of the corner resonator to its two immediate neighbors is identical, thus there is no difference between a sharp 90º bend and a linear straight waveguide for CDSW. This waveguiding mechanism is in great contrast with that of conventional spoof surface plasmons waveguide, where backscattering and radiation seriously impair the capability of these structures to bend surface waves with a small bending radius. Second, the waveguiding bands of this sharp CDSW corner lie in the photonic band gap of the gapped surface-wave photonic crystal, thus the coupled defect surface modes are tightly confined and propagate through this sharp corner, similar to the sharp bend in silicon photonic crystal[33]. Third, the waveguding mechanism of CDSW is via "hopping" due to the interaction between the neighboring evanescent cavity modes, which is similar to the coupled resonator optical waveguide (CROW)[6] or coupled-cavity waveguide (CCW)[7]. Note that these waveguiding mechanisms that guarantee the perfect transmission through a sharp corner are different with that of the recently proposed photonic topological insulators[34-37], which support backscattering-immune topological protected edge states arise from spin-orbit coupling.

Power splitters are important elements in realizing subwavelenth interconnections and routing of surface waves. Here we use this waveguiding mechanism to split surface waves into two arms, as illustrated in the inset of Fig. 5(a). Both the input and output waveguides contain six coupled cavities to construct a T-shaped splitter. As shown in Fig. 5(a), the propagating surface defect modes inside the input CDSW split almost equally into two CDSW output ports for all frequencies within the waveguiding band of surface defect modes. We also measure and simulate the electric field distributions on the T-shaped splitter, as shown in Fig. 5(b) and Fig. 5(c), respectively.

In conclusion, we have experimentally demonstrated a waveguiding mechanism to manipulate surface waves at a subwavelength scale. The guiding and bending of surface waves through localized surface defect cavities via near-field coupling are fundamentally different from the conventional spoof surface plasmon waveguides. High transmission of surface waves along a straight waveguide, around a sharp corner, as well as through a T-splitter are directly observed. These results provide opportunities to manipulate surface wave at a subwavelength scale with minimal crosstalk.

**Acknowledgements**

This work was sponsored by the NTU Start-Up Grants, Singapore Ministry of Education under Grant No. MOE2015-T2-1-070 and MOE2011-T3-1-005.

**Competing Financial Interests statement**

The authors declare no competing financial interests.

# References

1. E. Yablonovitch, Phys. Rev. Lett. **58**, 2059-2062 (1987).

2. S. John, Phys. Rev. Lett. **58**, 2486-2489 (1987).

3. J. D. Joannopoulos, R. D. Meade, and J. N. Winn, *Photonic Crystals: Molding the Flow of Light. Princeton*. (Princeton University Press, 1995.)

4. E. Yablonovitch, T. J. Gmitter, R. D. Meade, A. M. Rappe, K. D. Brommer, and J. D. Joannopoulos, Phys. Rev. Lett. **67**, 3380 (1991).

*5.* N. Stefanou and A. Modinos, Phys. Rev. B **57**, 12127 (1998).

6. A. Yariv, Y. Xu, R. K. Lee, and A. Scherer, Opt. Lett. **24**, 711 (1999).

7. M. Bayindir, B. Temelkuran, and E. Ozbay, Phys. Rev. Lett. **84**, 2140 (2000).

8. Y. Hara, T. Mukaiyama, K. Takeda, and M. Kuwata-Gonokami, Phys. Rev. Lett. **94**, 203905 (2005).

9. Shayan Mookherjea, Jung S. Park, Shun-Hui Yang, and Prabhakar R. Bandaru, Nature Photon. **2**, 90-93 (2008).

*10.* A. F. Harvey, IRE Trans. Microw. Theory Tech. **8**, 30-61 (1960).

*11.* B. A. Munk, *Frequency Selective Surfaces: Theory and Design.* (Wiley-Interscience, New York, 2000).

12. R. Collin, *Field Theory of Guided Waves.* (IEEE, New York, 1991).

13. J. B. Pendry, L. Martín-Moreno, and F. J. Garcia-Vidal, Science **305**, 847-848 (2004).

14. A. P. Hibbins, B. R. Evans, and J. R. Sambles, Science **308**, 670-672 (2005).

15. C. R. Williams, S. R. Andrews, S. A. Maier, A. I. Fernández-Domínguez, L. Martín-Moreno, and F. J. García-Vidal, Nat. Photonics **2**, 175-179 (2008).

16. D. Martin-Cano, M. L. Nesterov, A. I. Fernandez-Dominguez, F. J. Garcia-Vidal, L. Martin-Moreno, and Esteban Moreno, Opt. Express **18**, 754-764 (2010).

17. Z. Gao, X. F. Zhang, and L. F. Shen, J. Appl. Phys. **108**, 113104 (2010).

18. Z. Gao, L. F. Shen, and X. D. Zheng, IEEE Photon. Technol. Lett., **24**, 1343–1345, (2012).

19. X. P. Shen, T. J. Cui, D. Martin-Cano, and F. J. Garcia-Vidal, Proc. Natl. Acad. Sci. **110**, 40-45 (2013).

20. F. Gao, Z. Gao, X. Shi, Z. Yang, X. Lin, J. D. Joannopoulos, M. Soljacic, H. Chen, L. Lu, Y. Chong, and B. Zhang, arXive: **1504**.07809 (2015).

21. Z. Gao, F. Gao, Y. M. Zhang, X. H. Shi, Z. J. Yang, B. L. Zhang, Appl. Phys. Lett. **107**, 041118 (2015).

22. Z. Gao, F. Gao, Y. M. Zhang, B. L. Zhang, Appl. Phys. Lett. **107**, 051545 (2015).

23. Min Qiu, Opt. Express **13**, 7583-7588 (2005).

24. Juliette Plouin, Elodie Richalot, Odile Picon, Mathieu Carras, and Alfredo de Rossi, Opt. Express **14**, 9982-9987 (2006).

25. Z. Gao, L. Shen, J.-J. Wu, T.-J. Yang, and X. Zheng, Opt. Commun. **285**, 2076 (2012).

26. K. Kim, J. Kim, H. Y. Park, Y. Lee, S. Kim, S. Lee, and C. Kee, Opt. Express **22**, 4050-4058 (2014).

27. M. Carras and Alfredo De Rossi, Opt. Lett. **31**, 47-49 (2006).

28. L. Stabellini, M. Carras, A. D. Rossi, G. Bellanca, IEEE J Quantum Electron **44**, 905–910 (2008).

29. Seong-Han Kim, Sang Soon Oh, Kap-Joong Kim, Jae-Eun Kim, Hae Yong Park, Ortwin Hess, and Chul-Sik Kee, Phys. Rev. B **91**, 035116 (2015).

30. Fabrice Lemoult, Nadège Kaina, Mathias Fink, and Geoffroy Lerosey, Nature Physics **9**, 55-60 (2013).

31. S. J. Berry, T. Campbell, A. P. Hibbins, and J. R. Sambles, Appl. Phys. Lett. **100**, 101107 (2012).

32. N.W. Ashcroft and N. D. Mermin, *Solid State Physics*. (Sounders, Philadelphia, 1976).

33. A. Mekis, J. C. Chen, I. Kurland, S. Fan, P. R. Villeneuve, and J. D. Joannopoulos, Phys. Rev. Lett. **77**, 3787 (1996).

34. L. Lu, J. D. Joannopoulos, and M. Soljacic, Nature Photon. **8**, 821-829 (2014).

35. A. B. Khanikaev, S. H. Mousavi, W. Tse, M. Kargarian, A. H. MacDonald, and G. Shvets. Nature Mater. **12**, 233-239 (2012).

36. L. H. Wu and X. Hu, Phys. Rev. Lett. **114**, 223901 (2015).

37. T. Ma, A. B. Khanikaev, S. H. Mousavi, and G. Shvets, Phys. Rev. Lett. **114**, 127401 (2015).

**Figures and captions**

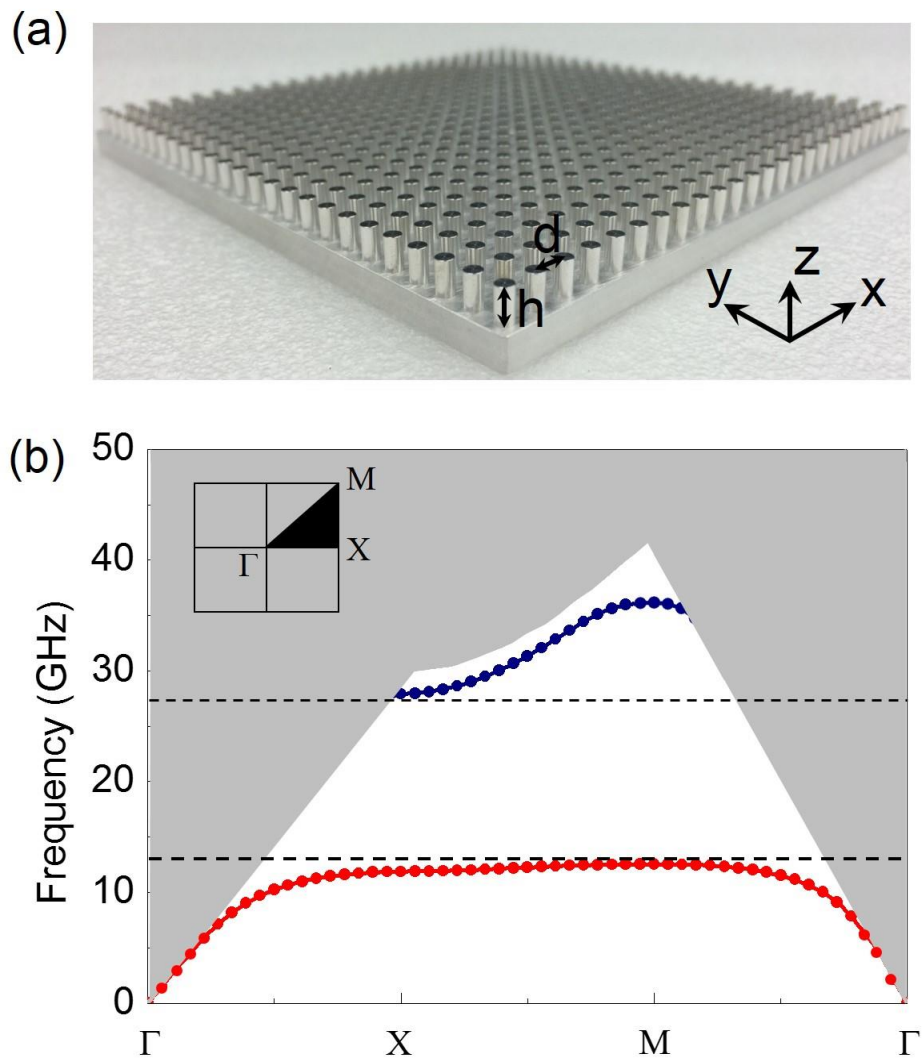

**FIG. 1.** (a) Photography of the surface-wave photonic crystal that consists of a square array of cylindrical pillars with radius $r = 1.25$ mm, height $h = 5$ mm, and periodicity $d = 5$ mm. (b) Projected photonic band structure of propagating surface modes. Radiation modes (shaded regions) above the light cone are not displayed.

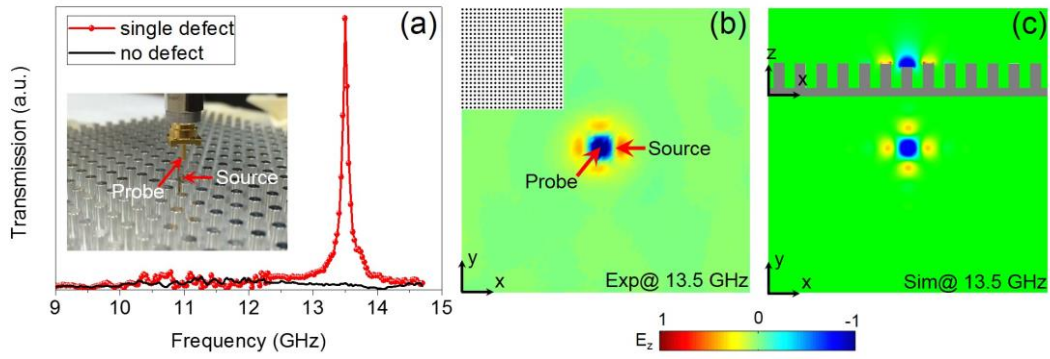

**FIG. 2.** (a) Measured near-field transmission spectra of a surface-wave photonic crystal with a single point defect (red line) and without any defect (black line). Inset shows the experimental set up. (b) Observed field pattern ($E_z$) when the source is inside the point defect at the resonance frequency 13.5 GHz. Inset shows the structured metallic surface with a point defect at the center. (c) Simulated field pattern ($E_z$) in the *xy* plane when the source is inside the point defect at the resonance frequency 13.5 GHz. Inset shows the vertical Ez field distribution in the central *xz* plane.

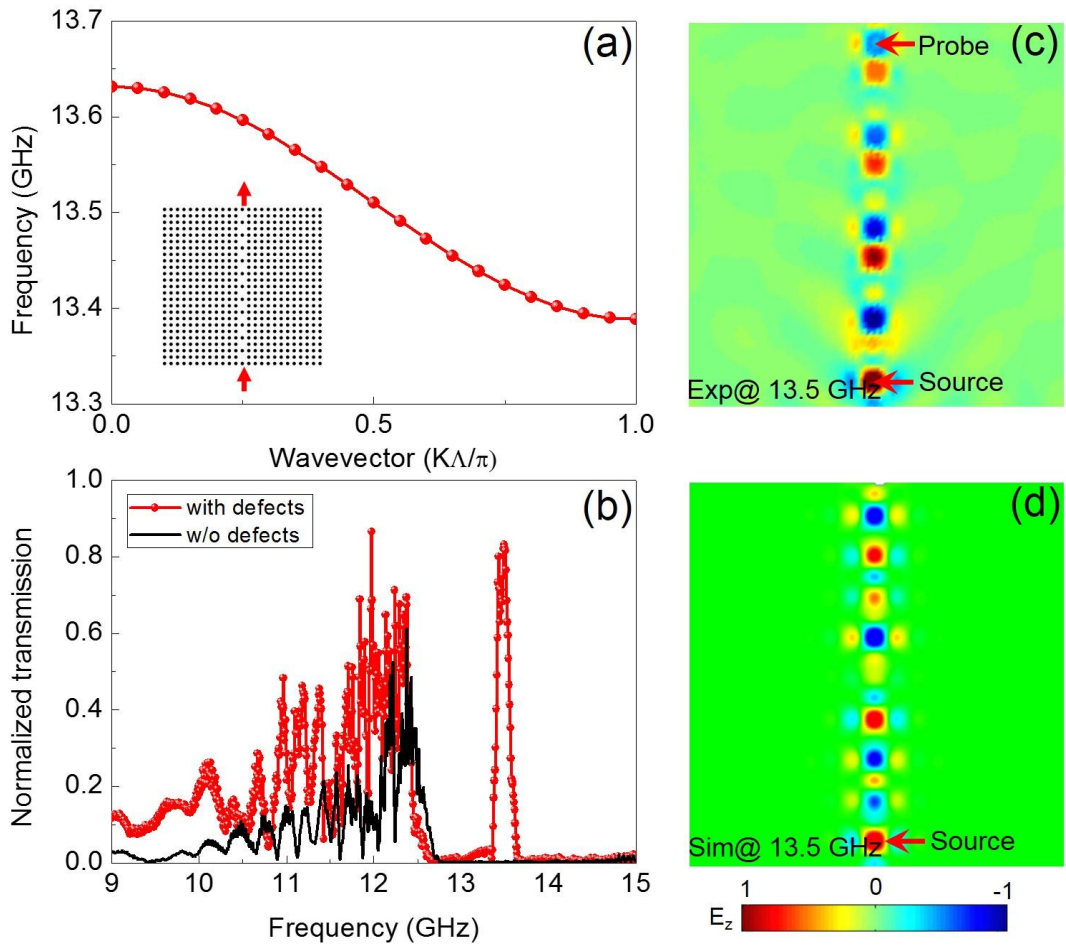

**FIG. 3.** (a) Simulated dispersion relation of the waveguiding band for the coupled defect surface waveguide. Inset illustrates the waveguide configuration. (b) Measured normalized transmission spectrum for a straight coupled defect surface waveguide (red line) and an ideal surface-wave photonic crystal without any defect (black line). (c) Measured field pattern ($E_z$) of a straight coupled defect surface waveguide at 13.5 GHz. (d) Simulated field pattern ($E_z$) of a straight coupled defect surface waveguide at 13.5 GHz.

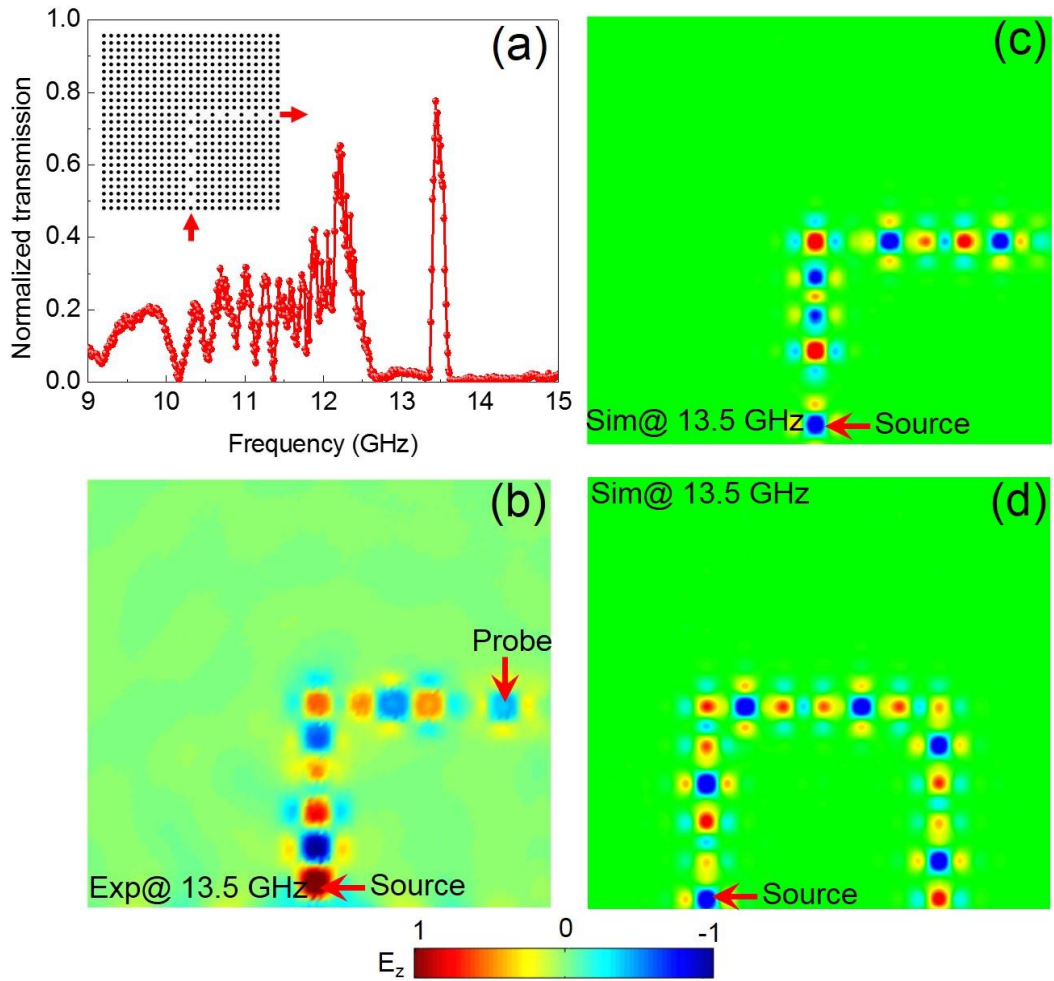

**FIG. 4.** (a) Measured normalized transmission spectrum of a 90-degree sharp corner. Inset shows the sharp corner geometry, where input/output ports are indicated with red arrows. (b) Observed field pattern ($E_z$) of a sharp corner at 13.5 GHz. Monopole source and probe are indicated with a pair of red arrows. (c) Simulated field pattern ($E_z$) of a sharp corner at 13.5 GHz. (d) Simulate field pattern ($E_z$) of a 180-degree bend which consists of two sharp corners.

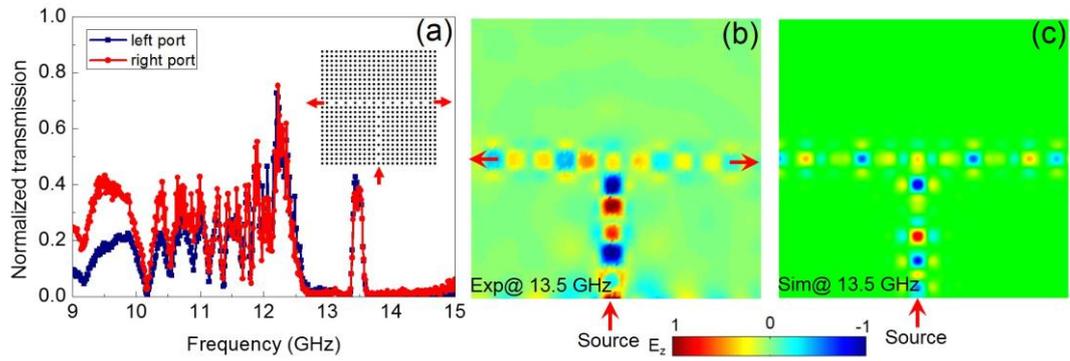

**FIG. 5.** (a) Measured normalized transmission spectra for the T-shaped splitter illustrated in the inset. Surface waves in the input waveguide split equally into two output waveguide. (b) Observed field pattern ($E_z$) of the T-shaped splitter at 13.5 GHz. Input/output ports are indicated with red arrows. (c) Simulated field pattern ($E_z$) of the T-shaped splitter at 13.5 GHz.